\documentclass[prb,preprint,superscriptaddress,showpacs]{revtex4}

\usepackage{epsfig} 
\usepackage{graphicx} 
\usepackage{amssymb, amstext, amsmath} 
\usepackage{bbold} 
\newcommand{\SOP}{2nd-PT}
\newcommand{\bs}{\hat} 	
\newcommand{\DeltaR}{\Delta_R}

\begin{document} 
\title{Accuracy of Second Order Perturbation theory in the Polaron and Variational Polaron Frames} 
\author{Chee Kong Lee} \email{cqtlck@nus.edu.sg} 
\affiliation{Centre for
Quantum Technologies, National University of Singapore, 117543, Singapore} 
\author{Jeremy  Moix} 
\affiliation{Department of Chemistry, Massachusetts
Institute of Technology, Cambridge, Massachusetts 02139, USA}
\affiliation{School of Materials Science and Engineering, Nanyang Technological
University, 639798, Singapore} 

\author{Jianshu Cao} 
\affiliation{Department of Chemistry, Massachusetts
Institute of Technology, Cambridge, Massachusetts 02139, USA}

\begin{abstract} 
In the study of open quantum systems, the polaron transformation has recently attracted a renewed interest 
   as it offers the possibility to explore the strong system-bath coupling regime.
   Despite this interest, a clear and unambiguous analysis of the
   regimes of validity of the polaron transformation is still lacking.
   Here we provide such a benchmark, comparing second
   order perturbation theory results in the original untransformed frame, the
   polaron
   frame and the variational extension with numerically exact 
   path integral calculations of the equilibrium reduced density matrix.
   Equilibrium quantities allow a direct comparison
   of the three methods without invoking any further approximations as is
   usually required in deriving master equations.
   It is found that the second order results in the original frame are accurate for
   weak system-bath coupling, the full polaron results are accurate in the 
   opposite regime of strong coupling, and the variational method is capable of 
   interpolating between these two extremes.
   As the bath becomes more non-Markovian (slow bath), all three approaches become 
   less accurate.
\end{abstract}
 \date{\today} \maketitle

%%%%%%%%%%%%%%%%%%%%%%%%%%%%%  INTRODUCTION  %%%%%%%%%%%%%%%%%%%%%%%
%%%%%%%%%%%%%%%%%%%%%%%%%%%%%  INTRODUCTION  %%%%%%%%%%%%%%%%%%%%%%%
%%%%%%%%%%%%%%%%%%%%%%%%%%%%%  INTRODUCTION  %%%%%%%%%%%%%%%%%%%%%%%
\section{Introduction} 

In many open quantum systems, the coupling between the system and the
bath can be considered as a small parameter.
In this case the application of second order perturbation leads 
to a master equation of Redfield or Lindblad type~\cite{breuer2002}. 
Their numerical implementation is straightforward and not computationally 
expensive. 
However, for many physical systems of current interest it has 
been shown that the weak coupling approximation is not justified. 
One example is the energy transfer process in photosynthetic complexes 
where the magnitude of the system-bath coupling is comparable to the 
electronic couplings~\cite{Vulto1998, Brixner2005, Cho2005, Wu2010}.  
There are only a few non-perturbative techniques to obtain the
numerically exact dynamics; examples include the hierarchy master
equation~\cite{Tanimura2006, Tanaka2009}, the quasi-adiabatic propagator
path integral (QUAPI)~\cite{Makri1995, Makri1995a}, and the multiconfiguration time-dependent 
Hartree (MCTDH)~\cite{Meyer1990, Beck2000, Thoss2001} approach.
However, these methods are computationally demanding and also not trivial
to implement. 

Recently, a polaron transformed second order master equation~\cite{Jang2008,
Jang2009, Jang2011, McCutcheon2010, McCutcheon2011b} and its
variational form~\cite{McCutcheon2011a, McCutcheon2011} have been 
derived to study the dynamics of open quantum systems at strong coupling. 
This approach transforms the total Hamiltonian into the polaron frame 
such that the system Hamiltonian is dressed by a polaron.
The master equation is then obtained by applying perturbation theory 
to the transformed system-bath interaction term. 
This approach extends the regime of validity of the master equation
to stronger system-bath coupling, provided that the 
electronic couplings (or tunneling matrix elements) are small compared 
to the typical bath frequency. 
When this condition is not fulfilled the polaron is too sluggish to 
accurately follow the system motion and the polaron transformation
may perform worse than the standard master equation approach.

In order to partially overcome this difficulty, the variational method has 
been developed as a generalization of the polaron transformation~\cite{Silbey1984, Harris1985}. 
Instead of performing the full transformation, the variational
polaron approach seeks for an optimal amount of transformation, 
depending on the properties of the bath. 
Thus it is able to interpolate between the strong and weak
coupling regimes and to capture the correct behavior
over a much broader range of parameters. 
Both the polaron and variational master equations have the attractive 
feature of being computationally economic
(they have the same computational complexity as the Redfield equation) 
and are therefore suitable for studying large systems. 

However, a thorough assessment of the accuracy of second order perturbation
theory in the polaron and variational polaron frames is still lacking. 
It is not exactly clear how the accuracy depends on the properties of the bath,
namely the bath relaxation time and the coupling strength. 
One of the main goals of this work is to provide such a benchmark.
Instead of studying the dynamics, here we focus on the equilibrium density 
matrix. 
Focusing on this quantity offers two key advantages.
Firstly, in the equilibrium case the second order perturbation is the only 
approximation involved.
In the derivation of second order master equations, 
additional approximations generally must be invoked, such as 
factorized initial conditions, 
the Born-Markov approximation, the rotating wave approximation, etc. 
These additional restrictions prevent a clear assessment of 
the isolated role of second order perturbation theory and
the merits of the polaron transformation.
Thus studying the equilibrium density matrix offers a 
direct comparison of the various perturbation methods.
A second advantage of studying the equilibrium state is that it is 
much easier to obtain numerically exact results. 
Therefore we are able to systematically explore a large range of
the parameter space that is often not possible with other exact treatments
of the dynamics.

%We derive the second order correction to the equilibrium state in both
%the polaron and variational polaron frames. 
%We then compare the results with numerically exact imaginary time path integral
%calculation (see Appendix). 
%Here we provide a direct comparison between the perturbation results and the
%numerically exact path integral calculation for the entire range of the bath
%parameters. 

In the next section, the details of the spin-boson model used in the
remainder of the text are outlined. 
Following this, the polaron transformation and its variational extension
are applied to the Hamiltonian in Sec.~\ref{sec:transformations}.
In Sec.~\ref{sec:pt}, the second-order corrections to the  
equilibrium reduced density matrix are derived in the original, polaron 
and variational polaron frames.
In the ensuing section, results for the various perturbation theories
are compared with exact numerical results from path integral calculations
over a broad range of the parameter space.
It is found that the second order results in the original frame are accurate for
small system-bath coupling, the full polaron results are accurate in the 
opposite regime of strong coupling, and the variational method is capable of 
interpolating between these two extremes.
All three approaches become less accurate for slow baths.

%%%%%%%%%%%%%%%%%%%%% SPIN BOSON %%%%%%%%%%%%%%%%%
%%%%%%%%%%%%%%%%%%%%% SPIN BOSON %%%%%%%%%%%%%%%%%	
%%%%%%%%%%%%%%%%%%%%% SPIN BOSON %%%%%%%%%%%%%%%%%

\section{Theory}
\subsection{Spin-Boson Model} \label{sec:spin-boson}

The spin-boson model is a paradigm for the study of quantum dissipative systems.
It has been used to investigate the energy transfer in light harvesting 
systems~\cite{pachon11,ishizaki09b}, 
the problem of decoherence in quantum optics~\cite{carmichael1999}, 
tunneling phenomena in condensed media~\cite{Leggett1987, weiss2008}, 
and quantum phase transitions~\cite{Chin2006, Chin2011}. 
The spin-boson model consists of a two-level system coupled to a
bath of harmonic oscillators.  Its Hamiltonian can be written as (we set $\hbar=1 $)
\begin{eqnarray}  %SPIN-BOSON HAMILTONIAN 
   \bs{H}'_{tot}=\frac{\epsilon}{2} \bs{\sigma}_z + \frac{\Delta}{2} \bs{\sigma}_x + \sum_k
   \omega_k \bs{b}^\dagger_k \bs{b}_k + \bs{\sigma}_z \sum_k g_k
   (\bs{b}^\dagger_k + \bs{b}_k), 
\end{eqnarray}
where $\sigma_i$ ($i=x, y, z$) are the usual Pauli matrices, $\epsilon$ is the
energy splitting between the two levels, and $\Delta$ is the tunneling matrix
element.
The bath is modeled as a set of harmonic oscillators 
labeled by their frequencies, $\omega_k$, and couplings to the two-level
system denoted by $g_k$.

The properties of the harmonic bath are completely determined by the spectral
density, $J(\omega)= \pi\sum_k g_k^2 \delta(\omega-\omega_k)$. Throughout the
paper, we use a super-ohmic spectral density with an exponential cut-off,
\begin{eqnarray}  %Spectral Density
	J(\omega) =\frac{1}{2}\gamma \frac{\omega^3}{\omega_c^3} 
        \mbox{e}^{-\omega/\omega_c},
\end{eqnarray}  
where $\gamma$ is the system-bath coupling strength and has the dimension of
frequency. The cut-off frequency is denoted by $\omega_c$, and 
its reciprocal governs the relaxation time of the 
bath, $\tau \propto \frac{1}{\omega_c}$.
%
%
%%%%%%%%%%%%%%%%%%%%%% TRANSFORMATION %%%%%%%%%%%%%%%%%%%%%
%%%%%%%%%%%%%%%%%%%%%% TRANSFORMATION %%%%%%%%%%%%%%%%%%%%%
%%%%%%%%%%%%%%%%%%%%%% TRANSFORMATION %%%%%%%%%%%%%%%%%%%%%

\subsection{Polaron and Variational Polaron Transformation}
\label{sec:transformations}

The polaron transformation is generated by the unitary operator 
\begin{eqnarray} % Unitary Transformation
		\bs{U}=\mbox{exp}\Big[\bs{\sigma}_z \sum_k \frac{f_k}{\omega_k}(\bs{b}_k^\dagger- \bs{b}_k)\Big],
\end{eqnarray}
which displaces the bath oscillators in the positive or negative direction 
depending on the state of the two-level system. 
The parameter $f_k$ determines the magnitude of the displacement for each mode.
Setting $f_k=g_k$ corresponds to the full polaron
transformation whereas $f_k=0$ corresponds to no transformation. 
The variational method allows us to determine an optimal value of $f_k$
that lies in between these two limits, $0\leq f_k \leq g_k$, 
making the transformation valid over a wider range of parameters.

Applying the transformation to the total Hamiltonian, we have  
\begin{eqnarray} % Transformed Hamiltonian
		\bs{H}_{tot} &=& \bs{U} \bs{H}'_{tot}\bs{U}^\dagger, \\
		             &=& \bs{H}_0 + \bs{H}_I, \nonumber
\end{eqnarray}
where the total free Hamiltonian is $\bs{H}_0=\bs{H}_S+\bs{H}_B$. 
The transformed system Hamiltonian is given by
\begin{eqnarray} % Transformed system Hamiltonian
 	\bs{H}_S=\frac{\epsilon}{2}\bs{\sigma}_z + \frac{\DeltaR}{2} \bs{\sigma}_x + \sum_k \frac{f_k }{\omega_k} (f_k-2g_k),\nonumber 
\end{eqnarray}
and the bath Hamiltonian remains unaffected, 
$\bs{H}_B=\sum_k \omega_k \bs{b}^\dagger_k \bs{b}_k$. 
The transformed interaction Hamiltonian becomes
\begin{eqnarray} %Coupling Hamiltonian 1
	 \bs{H}_I=\bs{\sigma}_x \bs{V}_x + \bs{\sigma}_y \bs{V}_y  +\bs{\sigma}_z \bs{V}_z,
\end{eqnarray}
where 
\begin{eqnarray} % Coupling Hamiltonian 2
	 \bs{V}_x &=& \frac{\Delta}{4} (\bs{D}_{+}^2 + \bs{D}_{-}^2 - 2B) ,\\
	 \bs{V}_y &=& \frac{\Delta}{4i} (\bs{D}_{-}^2 -\bs{D}_{+}^2) , \\
	 \bs{V}_z &=& \sum_k(g_k- f_k)(\bs{b}^\dagger_k + \bs{b}_k),
\end{eqnarray}
and $\bs{D}_{\pm}$ is the product of displacement operators,
$\bs{D}_{\pm}=\prod_k \mbox{e}^{\pm \frac{f_k}{\omega_k}(\bs{b}^\dagger_k
-\bs{b}_k)}$. 
The tunneling rate is renormalized by the expectation value of the
bath displacement operators, $\DeltaR=B \Delta$, where
\begin{eqnarray}
 	 B=\langle \bs{D}_{\pm}^2
\rangle_{\bs{H}_B} &=& \frac{\mbox{tr}_B[\bs{D}_{\pm}^2 \mbox{e}^{-\beta \bs{H}_B}]
}{\mbox{tr}_B[\mbox{e}^{-\beta \bs{H}_B}]}, \\
&=& \mbox{exp}\Big[-2\sum_k\frac{f_k^2}{\omega_k^2} \coth(\beta\omega_k /2)\Big].
\end{eqnarray}
Note that the interaction term is constructed such that 
$\langle \bs{H}_I \rangle_{H_0} = 0$.

Following Silbey and Harris~\cite{Silbey1984, Harris1985}, we use the
Bogoliubov variational theorem to determine the optimal values for the 
set $\{f_k\}$.
We first compute the Bogoliubov-Feynman upper bound on the free energy, $A_B$
\begin{eqnarray} % Free Energy
		A_B=-\frac{1}{\beta} \,\mbox{ln}\, \mbox{tr}_{S+B}\, [\mbox{e}^{-\beta \bs{H}_0}] + \langle \bs{H}_I \rangle_{\bs{H}_0}.
\end{eqnarray}
Since $\langle \bs{H}_I \rangle_{\bs{H}_0}=0$ by construction, the
upper bound is solely determined by the free Hamiltonian. The variational
theorem states that $A_B \geq A$ where $A$ is the true free energy. Therefore,
we want to make this bound as small as possible by minimizing $A_B$ with
respect to $\{f_k\}$, i.e. $\frac{d A_B}{d f_k}=0$. The minimization condition
leads to
\begin{eqnarray} % Minimization Results
		f_k &=& g_k F(\omega_k),\\
		F(\omega_k)  &=& \Big[1+\frac{\DeltaR^2}{\omega_k \eta} \coth(\beta \omega_k /2) \tanh(\beta\eta/2) \Big]^{-1}, \label{F-omega}
\end{eqnarray}
where $\eta=\sqrt{\epsilon^2 + \DeltaR^2}$.

In the continuum limit, the renormalization constant can be written as 
\begin{eqnarray} % Re-normalization constant
		B &=&  \mbox{exp}\Big[-2 \int^\infty_0 \frac{d\omega}{\pi}\frac{J(\omega)}{\omega^2} F(\omega)^2  \coth(\beta\omega /2)\Big]. \label{self-consistent}
\end{eqnarray}
Since $F(\omega)$ is also a function of $B$, the above equation must be solved self-consistently.

%%%%%%%%%%%%%%%%%%%%%%% 2ND ORDER PERTURBATION %%%%%%%%%%%%%
%%%%%%%%%%%%%%%%%%%%%%% 2ND ORDER PERTURBATION %%%%%%%%%%%%%
%%%%%%%%%%%%%%%%%%%%%%% 2ND ORDER PERTURBATION %%%%%%%%%%%%%
\subsection{Second Order Perturbation Theory}
\label{sec:pt}
This section is dedicated to finding the second order correction to the 
equilibrium state of the system. 
The exact equilibrium reduced density matrix can be formally
written as 
\begin{eqnarray}  % Exact equilibrium state
	\bs{\rho}_S=\frac{\mbox{tr}_B[\mbox{e}^{-\beta \bs{H}_{tot}}]}{\mbox{tr}_{S+B}[\mbox{e}^{-\beta \bs{H}_{tot}}]}. \label{exact_DM}
\end{eqnarray}
Expanding the operator $\mbox{e}^{-\beta \bs{H}_{tot} }$ up to second order in $\bs{H}_I$, we have
\begin{eqnarray} % Perturbative expansion
		\mbox{e}^{-\beta \bs{H}_{tot}} \approx \mbox{e}^{-\beta \bs{H}_0} \Big[1 -\int ^\beta_0 d \beta'  \mbox{e}^{\beta' \bs{H}_0} \bs{H}_I \mbox{e}^{-\beta' \bs{H}_0}
		  +  \int ^\beta_0 d \beta' \int ^{\beta'}_0 d \beta''  \mbox{e}^{\beta' \bs{H}_0} \bs{H}_I \mbox{e}^{-(\beta'- \beta'') \bs{H}_0} \bs{H}_I \mbox{e}^{-\beta'' \bs{H}_0}	\Big]. \nonumber \\
\end{eqnarray}
The above expansion is similar to the Dyson expansion, with $\beta$ treated as imaginary time. 

Since $\langle \bs{H}_I \rangle _{\bs{H}_0}=0$, the leading order correction to
$\bs{\rho}_S$ is of second order in $\bs{H}_I$. Inserting the above
expression into Eq.~(\ref{exact_DM}) and keeping terms up to the second order
in $\bs{H}_I$, the system equilibrium state can be approximated
as~\cite{Laird1991, Meier1999, Geva2000}
\begin{eqnarray}
	\bs{\rho}_S &\approx& \bs{\rho}_S^{(0)} + \bs{\rho}_S^{(2)} + \dots ;\\
		\bs{\rho}_S^{(0)} &=& \mbox{e}^{-\beta \bs{H}_S}/ Z_S^{(0)}; \nonumber \\
		\bs{\rho}_S^{(2)} &=& \frac{\bs{A}}{Z^{(0)}_S} - \frac{Z_S^{(2)}}{(Z^{(0)}_S)^2} \mbox{e}^{-\beta \bs{H}_S}, \label{second-order-state}\nonumber
\end{eqnarray}
where 
\begin{eqnarray} % Matrix A
		\bs{A} &=& \int^\beta _0 d\beta'  \int^{\beta'}_0 d\beta''  \sum_{nm} C_{nm}(\beta'-\beta'')
		  \times \mbox{e}^{(\beta' -\beta)\bs{H}_S} \bs{\sigma}_n  \mbox{e}^{(\beta'' -\beta')\bs{H}_S} \bs{\sigma}_m  \mbox{e}^{-\beta''\bs{H}_S}, \nonumber\\
		Z^{(0)}_S &=& \mbox{tr}_S[ \mbox{e}^{-\beta \bs{H}_S}] ,\,\,\,\,\,\, \, Z^{(2)}_S= \mbox{tr}_S[\bs{A}],
\end{eqnarray}
and $C_{nm}(\tau)$ is the bath correlation function in imaginary time, 
\begin{eqnarray} % Imaginary time correlation function
		C_{nm}(\tau)=\frac{\mbox{tr}_B[ \mbox{e}^ {\tau \bs{H}_B}\bs{V}_n \mbox{e}^ {-\tau \bs{H}_B}\bs{V}_m\mbox{e}^{-\beta \bs{H}_B}]}{\mbox{tr}_B[\mbox{e}^{-\beta \bs{H}_B}]}.
\end{eqnarray}

The non-vanishing bath correlation functions are 
\begin{eqnarray} % Bath Correlation Functions 
%(I removed the \hbar here since we set them to 1 earlier.)
	 %C_{zy}( \tau)&=& -C_{yz}( \tau) \\
	 							%&=& i \Delta_R \int^\infty_0 \frac{d\omega}{\pi} \frac{J(\omega)}{\omega}F(\omega)[1-F(\omega)] \frac{\sinh\Big(\frac{1}{2}(\beta -2\tau)\hbar \omega \Big)}{\sinh(\beta \hbar \omega/2)},\\
	 %C_{xx}( \tau) &=& \frac{\Delta_R^2}{8} (\mbox{e}^{\phi(\tau)} + \mbox{e}^{-\phi(\tau)} -2),\\
	 %C_{yy}( \tau) &=& \frac{\Delta_R^2}{8} (\mbox{e}^{\phi(\tau)} - \mbox{e}^{-\phi(\tau)} ),\\     
	 %C_{zz}( \tau) &=&  \int^\infty_0 \frac{d\omega}{\pi} J(\omega)[1-F(\omega)]^2 \frac{\cosh\Big(\frac{1}{2}(\beta -2\tau)\hbar \omega \Big)}{\sinh(\beta \hbar \omega/2)}, 
	 C_{zy}( \tau)&=& -C_{yz}( \tau) \\
	 							&=& i \Delta_R \int^\infty_0 \frac{d\omega}{\pi} \frac{J(\omega)}{\omega}F(\omega)[1-F(\omega)] \frac{\sinh\Big(\frac{1}{2}(\beta -2\tau) \omega \Big)}{\sinh(\beta \omega/2)},\\
	 C_{xx}( \tau) &=& \frac{\Delta_R^2}{8} (\mbox{e}^{\phi(\tau)} + \mbox{e}^{-\phi(\tau)} -2),\\
	 C_{yy}( \tau) &=& \frac{\Delta_R^2}{8} (\mbox{e}^{\phi(\tau)} - \mbox{e}^{-\phi(\tau)} ),\\     
	 C_{zz}( \tau) &=&  \int^\infty_0 \frac{d\omega}{\pi} J(\omega)[1-F(\omega)]^2 \frac{\cosh\Big(\frac{1}{2}(\beta -2\tau) \omega \Big)}{\sinh(\beta \omega/2)}, 
\label{eq:correlation_functions}
\end{eqnarray}
where
\begin{eqnarray} % \phi
		%\phi(\tau) = 4  \int^\infty_0 \frac{d\omega}{\pi} \frac{J(\omega)}{\omega^2} F(\omega)^2\frac{\cosh\Big(\frac{1}{2}(\beta -2\tau)\hbar \omega \Big)}{\sinh(\beta \hbar \omega/2)}.
		\phi(\tau) = 4  \int^\infty_0 \frac{d\omega}{\pi} \frac{J(\omega)}{\omega^2} F(\omega)^2\frac{\cosh\Big(\frac{1}{2}(\beta -2\tau) \omega \Big)}{\sinh(\beta  \omega/2)}.
\end{eqnarray}

It is useful at this point to analyze the behavior of
the perturbation theory at strong coupling in the polaron frame.
As seen from Eq.~(\ref{self-consistent}), when $\gamma\rightarrow\infty$ 
then $B\rightarrow 0$ and the system becomes incoherent
since the coherent tunneling element vanishes.
At the same time, $F(\omega)\rightarrow 1$ as $B \rightarrow 0$ 
so that all of the above correlation functions vanish, and hence also
the second order correction to $\hat \rho_S$.
Therefore in this limit, the equilibrium density matrix is only determined 
by the energy splitting of
the two levels, $\bs{\rho}_{S} \propto \mbox{exp}(-\frac{\epsilon}{2}\beta
\bs{\sigma}_z)$.

The full polaron result can be conveniently obtained by setting $F(\omega)=1$;
the only non-vanishing correlation functions in this case
are $C_{xx}$ and $C_{yy}$. 
The opposite limit of $F(\omega)=0$ corresponds to performing no 
transformation and $C_{zz}$ is the only non-zero correlation function. 
For comparison, below we will also include the results from these two limiting 
cases.

%%%%%%%%%%%%%%%%%%%%%%%%%%% RESULTS AND DISCUSSIONS %%%%%%%%%%%%%%%%
%%%%%%%%%%%%%%%%%%%%%%%%%%% RESULTS AND DISCUSSIONS %%%%%%%%%%%%%%%%
%%%%%%%%%%%%%%%%%%%%%%%%%%% RESULTS AND DISCUSSIONS %%%%%%%%%%%%%%%%
\section{Results and Discussions}
\label{sec:results}

In this section, we compare the results from second order perturbation theory (\SOP) in
the original [$F(\omega)=0$], the polaron [$F(\omega)=1$], and the 
variational polaron [$F(\omega)$ as in Eq.~(\ref{F-omega})] frames with those 
from numerically exact imaginary time path integral calculations. 
We compute the expectation value, $\langle \bs{\sigma}_z \rangle$, 
since it is not affected by the transformation, $\langle \bs{U}^\dagger \bs{\sigma}_z
\bs{U} \rangle = \langle \bs{\sigma}_z \rangle$. 
Therefore this quantity allows us to make a direct comparison between the 
path integral results, which provide the density matrix in the original frame, 
and the (variational) polaron results. 
Results from the transformed zeroth order density matrix, 
$\bs{\rho}^{(0)}_S$, which depends only on the renormalized system Hamiltonian $\bs{H}_S$, are also included.

We first calculate $\langle \bs{\sigma}_z \rangle$ as a function of the
dissipation strength for fast, slow and adiabatic baths, assessing the accuracy
of \SOP\ for different bath cut-off frequencies. 
We then conclude this section by presenting
phase diagrams of the relative errors of the various methods as functions
of the dissipation strength and the bath cut-off frequency.
This allows us to establish the regimes of validity of each approach
across the entire range of bath parameters. 
Throughout the paper, we set $\epsilon=1$ and $\beta=1$.\\

%%%%%%%%%%%%%%%%%%%% FAST BATH %%%%%%%%%%%%%%%
%%%%%%%%%%%%%%%%%%%% FAST BATH %%%%%%%%%%%%%%%
\subsection{Fast Bath, $\omega_c > \Delta$}

\begin{figure}  % FIGURE: PI versus Variational Results
   %\ifthenelse{\boolean{amCheeKong} }{
	 %\includegraphics[width=3.75in]{Figures/Var_Delta3_Omega5.pdf}
      %}{
	 \includegraphics[width=3.75in]{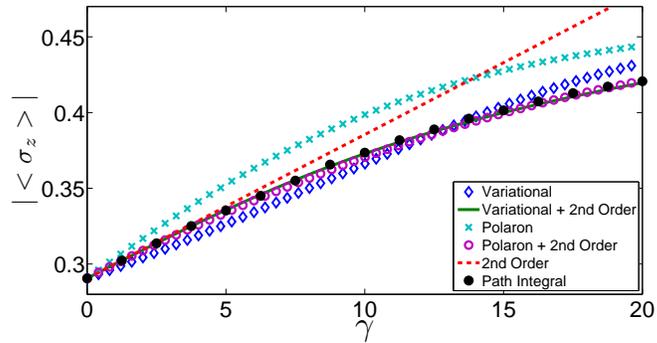}
      %}
    \caption{ (Color online) Fast bath, $\omega_c > \Delta$. 
    Comparison of the approximate methods with the exact path
    integral results as a function of the dissipation, $\gamma$, for
    $\omega_c=5\epsilon$, and $\Delta=3\epsilon$. 
    Plotted are the values of $\langle \bs{\sigma}_z \rangle$ from the 
    zeroth order density matrix in the polaron frame (light blue crosses) 
    and the variational polaron frame (dark blue diamonds), as
    well as from the density matrix in the original frame 
    with the second order correction (red dashed line), 
    the polaron frame (purple empty circles) 
    and the variational polaron frame (solid green line). 
    The exact path integral results are shown as filled (black) dots. 
    }\label{FIG:fast-bath}
\end{figure}

The value of $\langle \bs{\sigma}_z \rangle$ is plotted 
as a function of the dissipation strength, $\gamma$, 
in Fig.~\ref{FIG:fast-bath} for a fast bath,
$\omega_c > \Delta$. Firstly, it can be seen that the result from the usual
\SOP\ in the original frame (dashed line) is linearly dependent on $\gamma$.
While this approach is accurate at small $\gamma$, it
quickly degrades as the coupling increases.
On the other hand, the results from \SOP\ in the polaron (empty
circles) and the variational polaron (solid line) frames are in excellent
agreement with the exact path integral result (solid dots)
over the entire range of dissipation. 
The zeroth-order result for $\langle \bs{\sigma}_z \rangle$ 
in the the polaron frame (crosses) tends to overestimate the correction, 
whereas the variational frame result (diamonds)
provides at least a qualitatively correct description. 

%%%%%%%%%%%%%%%%%%%%%%%% SLOW BATH %%%%%%%%%%%%%%%%
%%%%%%%%%%%%%%%%%%%%%%%% SLOW BATH %%%%%%%%%%%%%%%%

\subsection{Slow Bath, $\omega_c < \Delta$}

\begin{figure}[h!]  % FIGURE:  Variational Results-- Small jump
   %\ifthenelse{\boolean{amCheeKong} }{
	 %\includegraphics[width=3.75in]{Figures/Var_Delta5_Omega15.pdf}
      %}{
	 \includegraphics[width=3.75in]{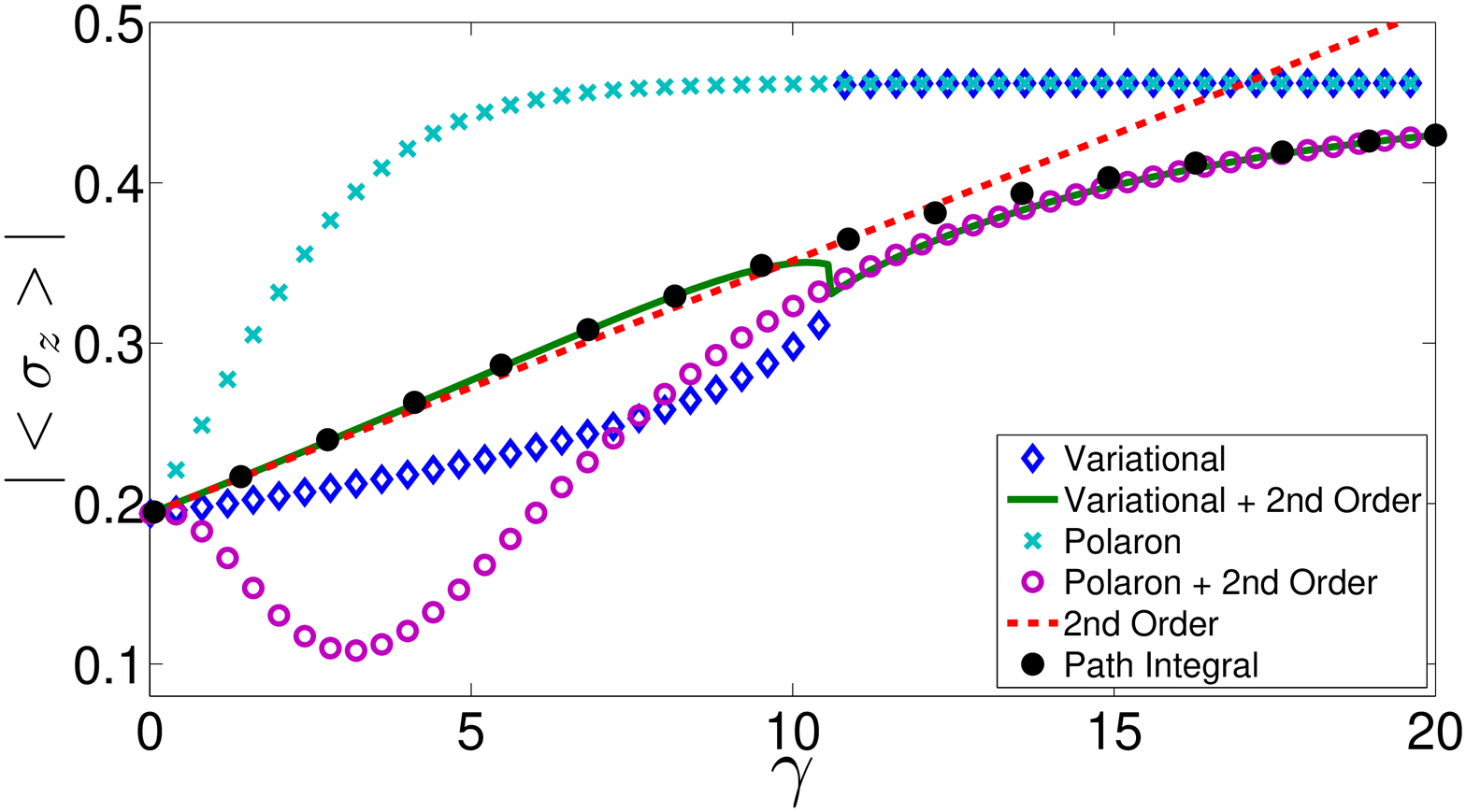}
      %}
    \caption{(Color online) Slow bath, $\omega_c < \Delta$. 
    Comparison of the approximate methods with the exact path
    integral results as a function of the dissipation, $\gamma$, for
    $\omega_c=1.5\epsilon$, and $\Delta=5\epsilon$. 
    Plotted are the values of $\langle \bs{\sigma}_z \rangle$ from the 
    zeroth order density matrix in the polaron frame (light blue crosses) 
    and the variational polaron frame (dark blue diamonds), as well as from
    the density matrix  in the original frame with the second order correction
    (red dashed line), the polaron frame (purple empty circles) 
    and the variational polaron frame (solid green line). 
    The exact path integral results are shown as filled (black) dots. 
    }\label{FIG:slow-bath}
\end{figure}

Fig.~\ref{FIG:slow-bath} displays the opposite case, when the bath is slow 
as compared to the tunneling rate, $\omega_c < \Delta$. 
At small and intermediate $\gamma$, it can be seen that the polaron method 
with \SOP\ fails to predict the correct behavior, while the usual \SOP\ result
in the original frame agrees well with the exact result. 
As with the case of the fast bath, at large $\gamma$, the polaron method
provides an accurate description, while the original frame \SOP\ breaks down.
The variational polaron method (with \SOP) interpolates between these two
methods, providing accurate results over a large range of the dissipation 
strength. 
The failure of the full polaron
method is due to the fact that the bath oscillators are sluggish and are not
able to fully dress the system. 
Therefore the full polaron displacement is no longer appropriate. 
It can also be seen that the second order correction in the 
full polaron frame is huge at small $\gamma$ (the
difference between the crosses and open circles). 
This should cast doubt on the validity of \SOP~in this case since
the perturbative correction should be small. 

\begin{figure}[h!]  % FIGURE:  Self-consistent  
   %\ifthenelse{\boolean{amCheeKong} }{
        %\includegraphics[width=3.75in]{Figures/self-consistent.pdf}
      %}{
        \includegraphics[width=3.75in]{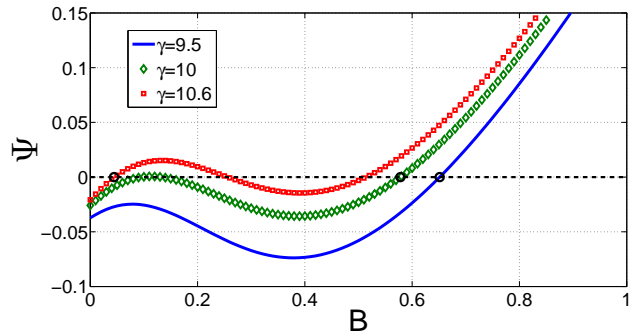}
      %}
     \caption{(Color online) 
     Plot of the expression $\Psi= B- \mbox{e}^{-2 \int
     \frac{d\omega}{\pi}\frac{J(\omega)}{\omega^2} F(\omega)^2
     \coth(\beta\omega /2)}$. The solutions to the self consistent equation
     Eq.~(\ref{self-consistent}) are the points when $\Psi=0$. 
     At $\gamma=9.5$ there is
     only one solution to the self-consistent equation. At $\gamma=10$,
     multiple solutions start to develop, but the solution with the lowest free
     energy (denoted by the empty circle) is chosen. At the critical point of
     $\gamma=10.6$, there is a ``jump'' in the lowest free energy solution
     which causes a discontinuity in the transformation. }
\label{FIG:self-consistent}
\end{figure}

It can also be observed that there is a discontinuity 
in the variational result (both with and without \SOP) 
at the critical point of $\gamma \approx 10.6$. 
The variational approach exhibits a rather abrupt transition from a small 
transformation [$F(\omega) \ll 1$] to the full polaron 
transformation [$F(\omega)\approx 1$].
This discontinuity, which is an artifact of the transformation rather than
a physical phase transition, has been predicted by 
Silbey and Harris~\cite{Silbey1984, Harris1985} for a slow bath. 
The discontinuity comes from solving the self-consistent equation in 
Eq.~(\ref{self-consistent}). Over a certain range of dissipation strengths,
there exist multiple solutions to the self-consistent equation. 
According to the variational prescription, the solution with the lowest 
free energy is selected. 
This causes a ``jump'' in the solution,
as depicted in Fig.~\ref{FIG:self-consistent}. 
It is also observed in Fig.~\ref{FIG:slow-bath} that the variational polaron 
result is least accurate around the transition point. 
Therefore, it will be worthwhile to look for a better variational criterion
that removes this discontinuity, which can hopefully provide a uniformly
accurate solution.

%\begin{figure}[h!]  % FIGURE: PI versus Variational Results
%   \includegraphics[width=3.5in]{Figures/Var_Delta3_Omega2.pdf}
%    \caption{$\Delta > \omega_c$. Comparison with exact path integral results as a function of dissipation, $\gamma$, for $\beta=1\epsilon$, cut-off frequency $\omega_c=2\epsilon$, and coupling $\Delta=3\epsilon$. Polaron transformation does work well at small dissipation.}
%\end{figure}

%%%%%%%%%%%%%%%%%%%%%   ADIABATIC BATH   %%%%%%%%%%%%%%%%%%%%%%%
%%%%%%%%%%%%%%%%%%%%%   ADIABATIC BATH   %%%%%%%%%%%%%%%%%%%%%%%
\subsection{Adiabatic Bath, $\omega_c \ll\beta ,\Delta$}
\begin{figure}[h!]  % FIGURE: Adiabatic Bath
	\center
   %\ifthenelse{\boolean{amCheeKong} }{
      %\includegraphics[width=3.75in]{Figures/Adiabatic-Bath.pdf}
   %}{
      \includegraphics[width=3.75in]{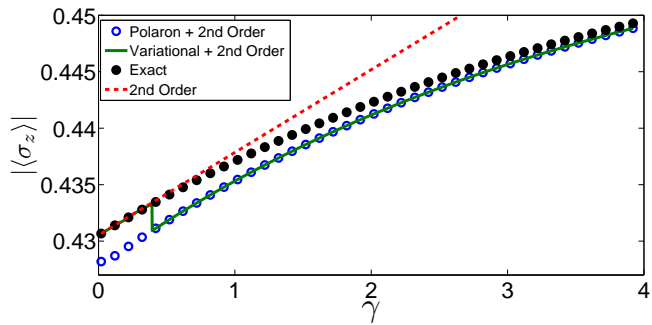}
   %}
    \caption{(Color online) $\omega_c \ll \Delta $. 
    Comparison of the approximate methods with the exact result [from
    Eq.~(\ref{adiabatic})] as a function of the dissipation, $\gamma$, for
    $\omega_c=0.1$, and $\Delta=1$. Plotted are the values of
    $\langle \bs{\sigma}_z \rangle$ from the density matrix with the 
    second order correction in the original frame (dashed red line), 
    the polaron frame (empty blue circles) and the
    variational polaron frame (solid green line). 
    The exact results are shown as filled (black) dots.}
    \label{FIG:adiabatic}
\end{figure}

In the adiabatic limit ($\omega_c \ll \beta $) the exact solution to the 
equilibrium state of the system can be obtained analytically. 
In this regime, the partition function is given by~\cite{chandler1991}
\begin{eqnarray} %Adiabatic Partition function
	 Z_S &=& \int^\infty_{-\infty} \frac{dx}{\sqrt{2\pi \chi}}\mbox{exp}\Big\{- \frac{x^2}{2 \chi}+ \ln\Big[2 \cosh[\beta \sqrt{(J/2)^2 +(x+\epsilon/2)^2}]\Big]\Big\},  \label{adiabatic}
\end{eqnarray}
where $\chi$ is the bath correlation function 
[in the original frame with $F(\omega)=0$] in
the adiabatic limit, $\chi= C_{zz}^{(0)} =\frac{2\gamma }{\pi \beta}$. 
The expectation value, $\langle \bs{\sigma}_z \rangle$, can be obtained 
from the partition function via the following relation
\begin{eqnarray} % SIGMA_Z from partition function
	 \langle\bs{\sigma}_z\rangle =-\frac{2}{\beta} \frac{\partial}{\partial\epsilon}  \mbox{ln} Z_S.
\end{eqnarray}

In the regime where $\omega_c \ll \Delta$, the transition from $F(\omega)=0$ to
$F(\omega)=1$ in the variational method is sharp, 
as seen in Fig.~\ref{FIG:adiabatic}.
Before the transition, the variational polaron result coincides with the exact
result and that of perturbation theory in the original frame. 
The full polaron result fails to give the correct results, and
even predicts the wrong limiting behavior as $\gamma \rightarrow 0$.
After the transition, the variational result deviates from the exact 
result and becomes essentially the same as the full polaron result. 
As $\gamma$ increases, results from both methods approach
the exact result while the untransformed \SOP\ breaks down as seen before.

%%%%%%%%%%%%%%%%%%%%%  RELATIVE ERROR  %%%%%%%%%%%%%%
%%%%%%%%%%%%%%%%%%%%%  RELATIVE ERROR  %%%%%%%%%%%%%%
\subsection{Relative Errors}
\begin{figure}[h!]  % FIGURE: CONTOUR PLOTS
  \center
  %\ifthenelse{\boolean{amCheeKong}}{
      %\includegraphics[width=3in]{Figures/2nd_contour.pdf}
      %\includegraphics[width=3in]{Figures/polaron_contour.pdf}
      %\includegraphics[width=3in]{Figures/variational_contour.pdf}
   %}{
      %\includegraphics[width=3in]{Figures/2nd_contour.eps}
      %\includegraphics[width=3in]{Figures/polaron_contour.eps}
      %\includegraphics[width=3in, bb=70 425 600 725]{tmp/figures.eps}
      \includegraphics[width=3in]{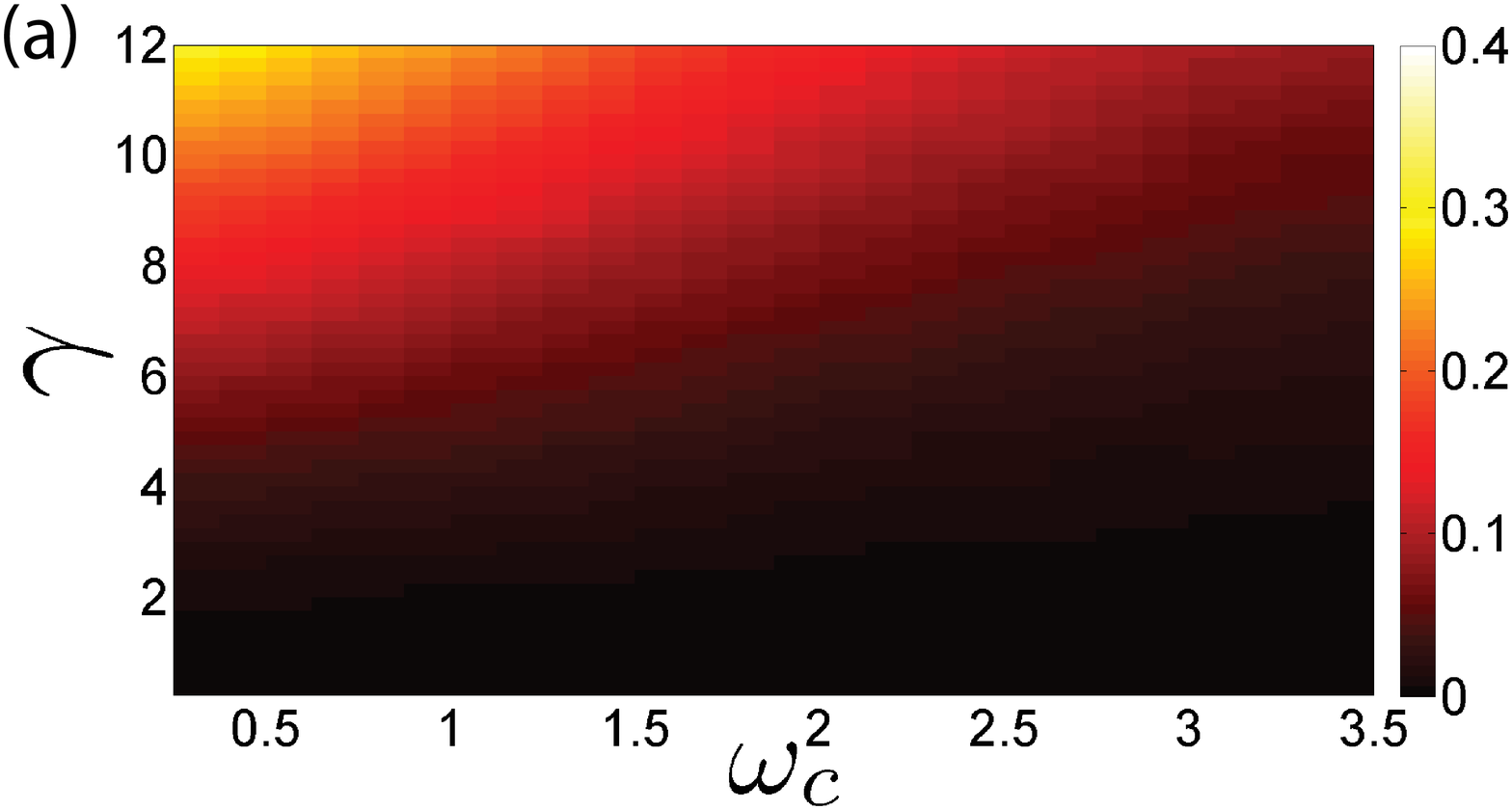}
      \includegraphics[width=3in]{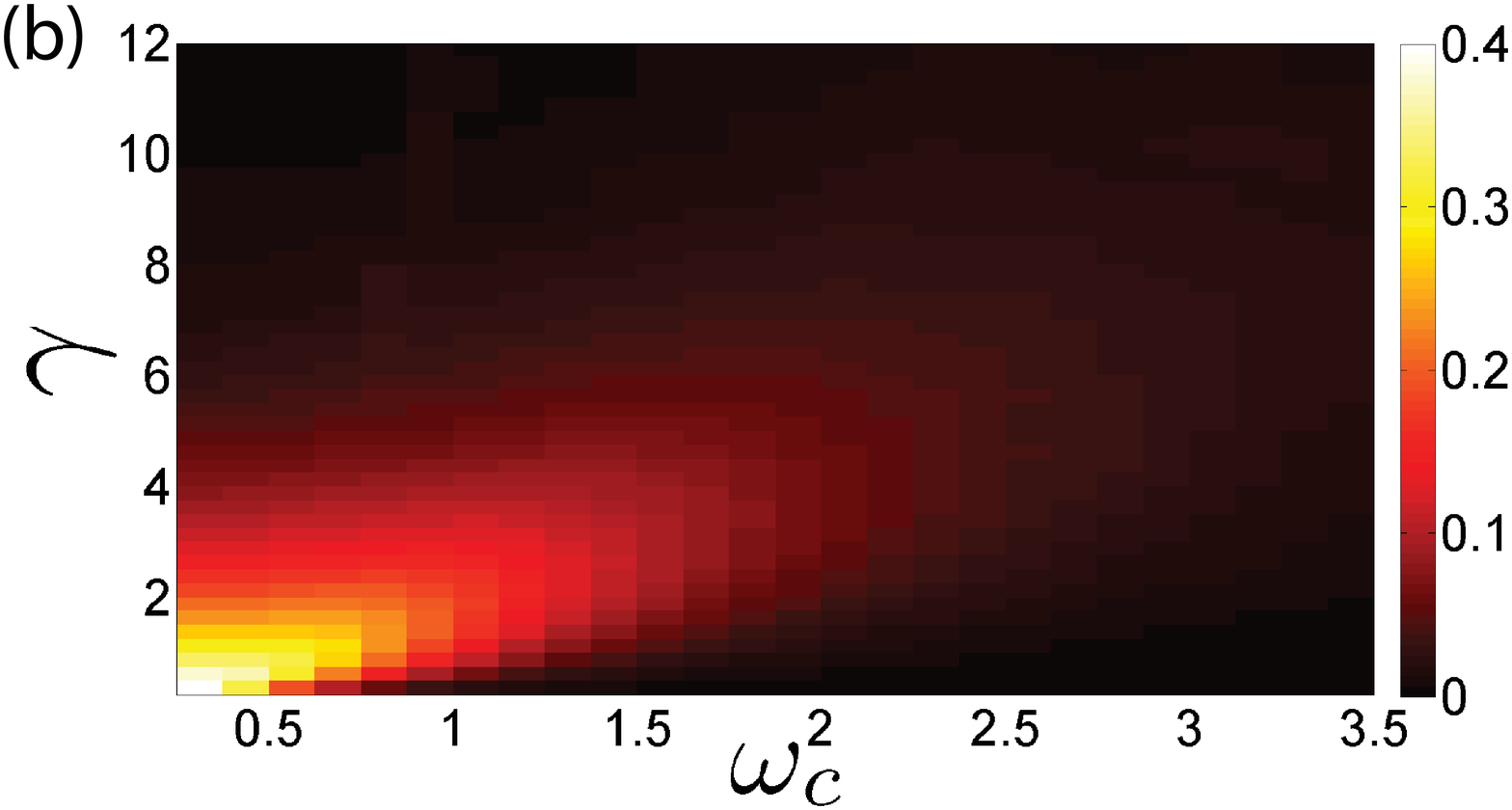}
      \includegraphics[width=3in]{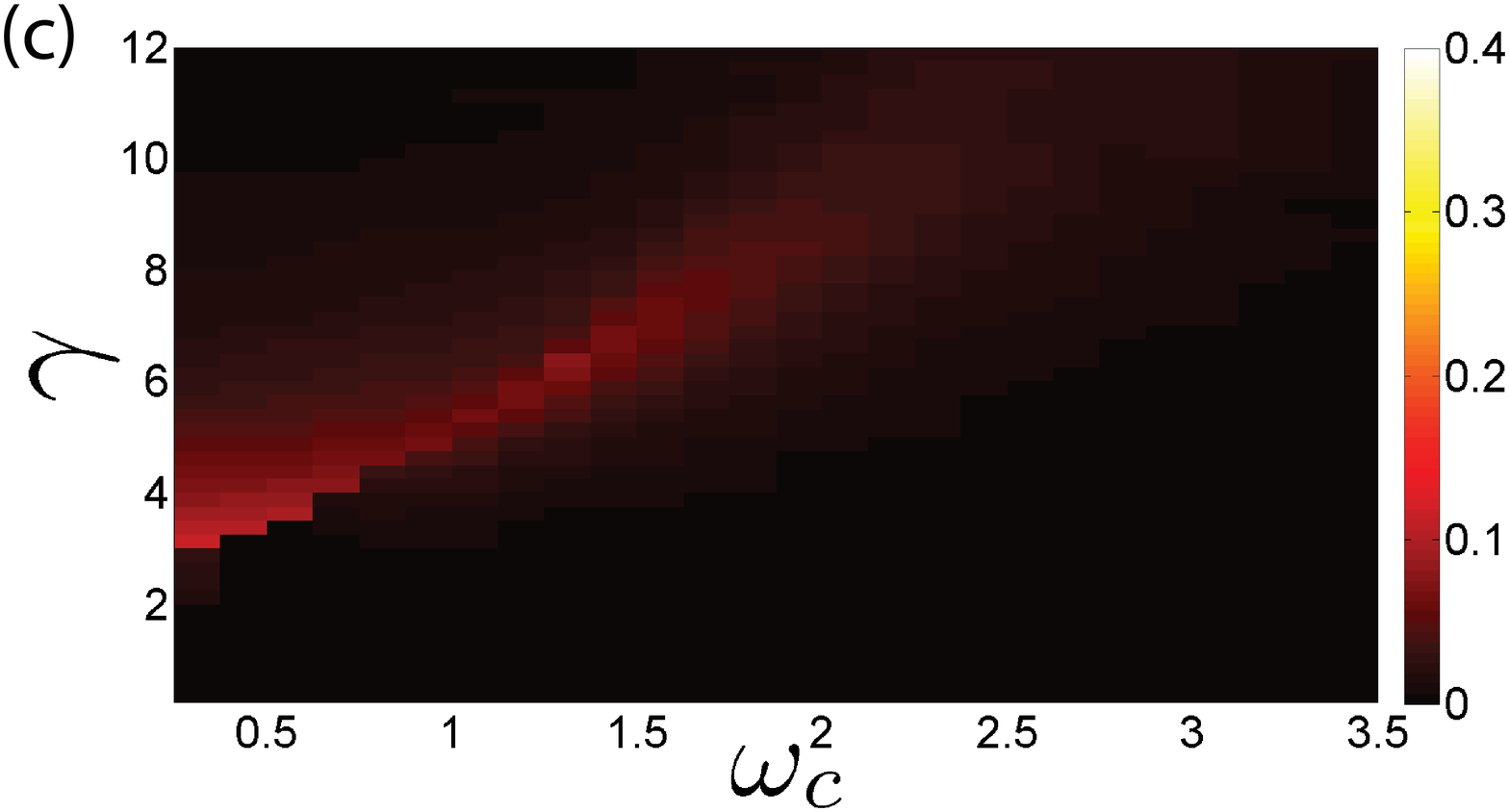}
   %}
         \caption{(Color online) At $\Delta=3$, the relative errors of the
         second order perturbation theory as defined in Eq.~(\ref{eq:error})
         in (a) the original frame, (b) the
         full polaron frame, and (c) the variational polaron frame.}
         \label{FIG:contour}
\end{figure}

To get a better perspective of how the accuracy of \SOP\ in different frames
depends on the properties of the bath, we calculate the relative errors over the entire range of the bath parameters. The relatives errors are defined as
\begin{eqnarray} % relative error
	\left|\frac{\langle \bs{\sigma}_z \rangle_{Pert}-\langle \bs{\sigma}_z \rangle_{PI}}{\langle \bs{\sigma}_z \rangle_{PI}}\right| \;,
        \label{eq:error}
\end{eqnarray}
where the subscripts ``Pert'' and ``PI'' denote the perturbative calculation and path integral calculation respectively. Fig.~\ref{FIG:contour} displays the respective errors for the three
methods as a function of the cut-off frequency and the coupling strength.
As seen in Fig.~\ref{FIG:contour}(a), the usual \SOP\ without 
transformation breaks down at large $\gamma$. 
It is also less accurate when the cut-off frequency is
small, which corresponds to a highly non-Markovian bath. 
On the other hand, the \SOP\ in the full polaron frame fails at 
small $\gamma$ and $\omega_c$ [see Fig.~\ref{FIG:contour}(b)]. 
These two approaches provide complementary behavior as a function of the 
coupling strength; the polaron method is essentially exact for large $\gamma$,
while the usual \SOP\ is exact for small $\gamma$.
The variational calculation is valid
over a much broader range of parameters [see Fig.~\ref{FIG:contour}(c)],
and essentially combines the regimes of validity of the full 
polaron result and \SOP~in the original frame. 
It is only slightly less accurate in the slow bath regime around the region
where the discontinuity appears that was discussed above.

%%%%%%%%%%%%%%%%%%%%%%%%%%%% CONCLUSIONS %%%%%%%%%%%%%%%%%%%%
%%%%%%%%%%%%%%%%%%%%%%%%%%%% CONCLUSIONS %%%%%%%%%%%%%%%%%%%%
%%%%%%%%%%%%%%%%%%%%%%%%%%%% CONCLUSIONS %%%%%%%%%%%%%%%%%%%%

\section{Conclusions}

In conclusion, we have provided a thorough assessment of the accuracy of 
the polaron and variational polaron methods.
We compared the second order perturbation results in the polaron and 
variational polaron
transformed frames with numerically exact path integral calculations of the
equilibrium reduced density matrix. 
Focusing on the equilibrium properties allowed us to 
systematically explore the whole range of bath parameters without
making any additional approximations as is generally required to simulate the
dynamics. 
As a function of the system-bath coupling, 
it is found that the standard perturbation result without the polaron 
transformation is accurate for small coupling, while
the polaron result is accurate in the opposite regime of strong coupling. 
The variational method is capable of interpolating between these two limits. 
It is valid over a much broader range of parameters 
and is only slightly less accurate around the region where the 
discontinuity appears.
As the relaxation time of the bath becomes longer leading to 
more non-Markovian character,
all three of the perturbation methods are seen to be less accurate.

%%%%%%%%%%%%%%%%%%%%%%%%%%%% ACKNOWLEDGEMENT %%%%%%%%%%%%%%%%
%%%%%%%%%%%%%%%%%%%%%%%%%%%% ACKNOWLEDGEMENT %%%%%%%%%%%%%%%%
\begin{acknowledgments}
The Centre for Quantum Technologies is a Research Centre of Excellence funded by the Ministry of Education 
and the National Research Foundation of Singapore.
J. Moix and J. Cao acknowledge support from NSF (Grant No. CHE-1112825) and DARPA (Grant No. N99001-10-1-4063).
J. Cao is supported as part of the Center for Excitonics, 
an Energy Frontier Research Center funded by the US Department of Energy,
Office of Science, Office of Basic Energy Sciences under Award No. DE-SC0001088. 
C. K. Lee is grateful to Leong Chuan Kwek and Jiangbin Gong for reading the manuscript and providing useful comments. 
\end{acknowledgments}

%%%%%%%%%%%%  APPENDIX  %%%%%%%%%%%%%
%%%%%%%%%%%%  APPENDIX  %%%%%%%%%%%%%
\appendix

\section{Imaginary Time Path Integral}

For Hamiltonians such as the spin-boson model where the bath is harmonic,
the trace over the bath degrees of freedom in Eq.~(\ref{exact_DM})
may be performed analytically.
In the path integral formulation this procedure leads to the well-known 
Feynman-Vernon influence 
functional.\cite{Feynman63,schul86,chandler1991,weiss2008,Ingold88,kleinert04}
Using a Hubbard-Stratonovich 
transformation,\cite{schul86,chandler1991,weiss2008,jeremy2012} 
it was shown that the influence functional 
may be unraveled by an auxiliary stochastic field.
The ensuing imaginary time evolution may then be interpreted as 
one governed by a time-dependent Hamiltonian.
Explicitly, the spin-boson model may be equivalently expressed as
\begin{eqnarray}  
   \bs{H}(\tau) = \frac{\epsilon}{2} \bs{\sigma}_z 
           + \frac{\Delta}{2} \bs{\sigma}_x 
           + \bs{\sigma}_z \xi(\tau).
\end{eqnarray}
All of the effects of the bath are accounted for by the 
colored noise term, $\xi(\tau)$, which obeys the autocorrelation relation,
\begin{equation}
   \left \langle \xi(\tau) \xi(\tau^{\prime})\right \rangle = 
    C_{zz}^{(0)}(\tau-\tau^{\prime}) \;,
\end{equation}
where $C_{zz}^{(0)}(\tau)$ is the correlation function
given in Eq.~(\ref{eq:correlation_functions}) with $F(\omega)=0$.
The trace over the bath that was present in the original path
integral formulation now corresponds to averaging the imaginary time
dynamics over realizations of the noise.
The auxiliary field is simply an efficient method of sampling the 
influence functional.

In practice, a sample of the reduced density matrix is propagated to 
the imaginary time $\beta$, where the time steps, $\delta \tau$,
are determined by
\begin{equation}
   \hat \rho_S(\tau + \delta \tau) = 
       \exp\left[-\delta\tau \bs{H}(\tau) \right] 
       \hat \rho_S(\tau)
       \;, \label{eq:schrodinger}
\end{equation}
with the initial condition, $\rho(0) = I$.
The primary benefit of this approach is that it generates the 
entire reduced density matrix from a single Monte Carlo calculation.
Additionally, any form for the spectral density of the bath, $J(\omega)$,
may be used. In our calculations, $10^8$ (at small $\gamma$) to $10^{11}$ (at large $\gamma$) Monte Carlo samples are needed to achieve convergence.

%\bibliography{myref-variational}

\begin{thebibliography}{37}
\expandafter\ifx\csname natexlab\endcsname\relax\def\natexlab#1{#1}\fi
\expandafter\ifx\csname bibnamefont\endcsname\relax
  \def\bibnamefont#1{#1}\fi
\expandafter\ifx\csname bibfnamefont\endcsname\relax
  \def\bibfnamefont#1{#1}\fi
\expandafter\ifx\csname citenamefont\endcsname\relax
  \def\citenamefont#1{#1}\fi
\expandafter\ifx\csname url\endcsname\relax
  \def\url#1{\texttt{#1}}\fi
\expandafter\ifx\csname urlprefix\endcsname\relax\def\urlprefix{URL }\fi
\providecommand{\bibinfo}[2]{#2}
\providecommand{\eprint}[2][]{\url{#2}}

\bibitem[{\citenamefont{Breuer and Petruccione}(2002)}]{breuer2002}
\bibinfo{author}{\bibfnamefont{H.-P.} \bibnamefont{Breuer}} \bibnamefont{and}
  \bibinfo{author}{\bibfnamefont{F.}~\bibnamefont{Petruccione}},
  \emph{\bibinfo{title}{The Theory of Open Quantum Systems}}
  (\bibinfo{publisher}{Oxford Univ. Press}, \bibinfo{year}{2002}).

\bibitem[{\citenamefont{Vulto et~al.}(1998)\citenamefont{Vulto, de~Baat, Louwe,
  Permentier, Neef, Miller, van Amerongen, and Aartsma}}]{Vulto1998}
\bibinfo{author}{\bibfnamefont{S.~I.~E.} \bibnamefont{Vulto}},
  \bibinfo{author}{\bibfnamefont{M.~A.} \bibnamefont{de~Baat}},
  \bibinfo{author}{\bibfnamefont{R.~J.~W.} \bibnamefont{Louwe}},
  \bibinfo{author}{\bibfnamefont{H.~P.} \bibnamefont{Permentier}},
  \bibinfo{author}{\bibfnamefont{T.}~\bibnamefont{Neef}},
  \bibinfo{author}{\bibfnamefont{M.}~\bibnamefont{Miller}},
  \bibinfo{author}{\bibfnamefont{H.}~\bibnamefont{van Amerongen}},
  \bibnamefont{and} \bibinfo{author}{\bibfnamefont{T.~J.}
  \bibnamefont{Aartsma}}, \bibinfo{journal}{J. Phys. Chem. B}
  \textbf{\bibinfo{volume}{102}}, \bibinfo{pages}{9577} (\bibinfo{year}{1998}).

\bibitem[{\citenamefont{Brixner et~al.}(2005)\citenamefont{Brixner, Stenger,
  Vaswani, Cho, Blankenship, and Fleming}}]{Brixner2005}
\bibinfo{author}{\bibfnamefont{T.}~\bibnamefont{Brixner}},
  \bibinfo{author}{\bibfnamefont{J.}~\bibnamefont{Stenger}},
  \bibinfo{author}{\bibfnamefont{H.~M.} \bibnamefont{Vaswani}},
  \bibinfo{author}{\bibfnamefont{M.}~\bibnamefont{Cho}},
  \bibinfo{author}{\bibfnamefont{R.~E.} \bibnamefont{Blankenship}},
  \bibnamefont{and} \bibinfo{author}{\bibfnamefont{G.~R.}
  \bibnamefont{Fleming}}, \bibinfo{journal}{Nature}
  \textbf{\bibinfo{volume}{434}}, \bibinfo{pages}{625} (\bibinfo{year}{2005}).

\bibitem[{\citenamefont{Cho et~al.}(2005)\citenamefont{Cho, Vaswani, Brixner,
  Stenger, and Fleming}}]{Cho2005}
\bibinfo{author}{\bibfnamefont{M.}~\bibnamefont{Cho}},
  \bibinfo{author}{\bibfnamefont{H.~M.} \bibnamefont{Vaswani}},
  \bibinfo{author}{\bibfnamefont{T.}~\bibnamefont{Brixner}},
  \bibinfo{author}{\bibfnamefont{J.}~\bibnamefont{Stenger}}, \bibnamefont{and}
  \bibinfo{author}{\bibfnamefont{G.~R.} \bibnamefont{Fleming}},
  \bibinfo{journal}{J. Phys. Chem. B} \textbf{\bibinfo{volume}{109}},
  \bibinfo{pages}{10542} (\bibinfo{year}{2005}).

\bibitem[{\citenamefont{Wu et~al.}(2010)\citenamefont{Wu, Liu, Shen, Cao, and
  Silbey}}]{Wu2010}
\bibinfo{author}{\bibfnamefont{J.}~\bibnamefont{Wu}},
  \bibinfo{author}{\bibfnamefont{F.}~\bibnamefont{Liu}},
  \bibinfo{author}{\bibfnamefont{Y.}~\bibnamefont{Shen}},
  \bibinfo{author}{\bibfnamefont{J.}~\bibnamefont{Cao}}, \bibnamefont{and}
  \bibinfo{author}{\bibfnamefont{R.~J.} \bibnamefont{Silbey}},
  \bibinfo{journal}{New J. Phys.} \textbf{\bibinfo{volume}{12}},
  \bibinfo{pages}{105012} (\bibinfo{year}{2010}).

\bibitem[{\citenamefont{Tanimura}(2006)}]{Tanimura2006}
\bibinfo{author}{\bibfnamefont{Y.}~\bibnamefont{Tanimura}},
  \bibinfo{journal}{J. Phys. Soc. Jpn.} \textbf{\bibinfo{volume}{75}},
  \bibinfo{pages}{082001} (\bibinfo{year}{2006}).

\bibitem[{\citenamefont{Tanaka and Tanimura}(2009)}]{Tanaka2009}
\bibinfo{author}{\bibfnamefont{M.}~\bibnamefont{Tanaka}} \bibnamefont{and}
  \bibinfo{author}{\bibfnamefont{Y.}~\bibnamefont{Tanimura}},
  \bibinfo{journal}{J. Phys. Soc. Jpn.} \textbf{\bibinfo{volume}{78}},
  \bibinfo{pages}{073802} (\bibinfo{year}{2009}).

\bibitem[{\citenamefont{Makri and Makarov}(1995{\natexlab{a}})}]{Makri1995}
\bibinfo{author}{\bibfnamefont{N.}~\bibnamefont{Makri}} \bibnamefont{and}
  \bibinfo{author}{\bibfnamefont{D.~E.} \bibnamefont{Makarov}},
  \bibinfo{journal}{J. Chem. Phys.} \textbf{\bibinfo{volume}{102}},
  \bibinfo{pages}{4600} (\bibinfo{year}{1995}{\natexlab{a}}).

\bibitem[{\citenamefont{Makri and Makarov}(1995{\natexlab{b}})}]{Makri1995a}
\bibinfo{author}{\bibfnamefont{N.}~\bibnamefont{Makri}} \bibnamefont{and}
  \bibinfo{author}{\bibfnamefont{D.~E.} \bibnamefont{Makarov}},
  \bibinfo{journal}{J. Chem. Phys.} \textbf{\bibinfo{volume}{102}},
  \bibinfo{pages}{4611} (\bibinfo{year}{1995}{\natexlab{b}}).

\bibitem[{\citenamefont{Meyer et~al.}(1990)\citenamefont{Meyer, Manthe, and
  Cederbaum}}]{Meyer1990}
\bibinfo{author}{\bibfnamefont{H.-D.} \bibnamefont{Meyer}},
  \bibinfo{author}{\bibfnamefont{U.}~\bibnamefont{Manthe}}, \bibnamefont{and}
  \bibinfo{author}{\bibfnamefont{L.}~\bibnamefont{Cederbaum}},
  \bibinfo{journal}{Chem. Phys. Lett.} \textbf{\bibinfo{volume}{165}},
  \bibinfo{pages}{73 } (\bibinfo{year}{1990}).

\bibitem[{\citenamefont{Beck et~al.}(2000)\citenamefont{Beck, Jäckle, Worth,
  and Meyer}}]{Beck2000}
\bibinfo{author}{\bibfnamefont{M.}~\bibnamefont{Beck}},
  \bibinfo{author}{\bibfnamefont{A.}~\bibnamefont{Jäckle}},
  \bibinfo{author}{\bibfnamefont{G.}~\bibnamefont{Worth}}, \bibnamefont{and}
  \bibinfo{author}{\bibfnamefont{H.-D.} \bibnamefont{Meyer}},
  \bibinfo{journal}{Phys. Rep.} \textbf{\bibinfo{volume}{324}},
  \bibinfo{pages}{1 } (\bibinfo{year}{2000}).

\bibitem[{\citenamefont{Thoss et~al.}(2001)\citenamefont{Thoss, Wang, and
  Miller}}]{Thoss2001}
\bibinfo{author}{\bibfnamefont{M.}~\bibnamefont{Thoss}},
  \bibinfo{author}{\bibfnamefont{H.}~\bibnamefont{Wang}}, \bibnamefont{and}
  \bibinfo{author}{\bibfnamefont{W.~H.} \bibnamefont{Miller}},
  \bibinfo{journal}{J. Chem. Phys.} \textbf{\bibinfo{volume}{115}},
  \bibinfo{pages}{2991} (\bibinfo{year}{2001}).

\bibitem[{\citenamefont{Jang et~al.}(2008)\citenamefont{Jang, Cheng, Reichman,
  and Eaves}}]{Jang2008}
\bibinfo{author}{\bibfnamefont{S.}~\bibnamefont{Jang}},
  \bibinfo{author}{\bibfnamefont{Y.-C.} \bibnamefont{Cheng}},
  \bibinfo{author}{\bibfnamefont{D.~R.} \bibnamefont{Reichman}},
  \bibnamefont{and} \bibinfo{author}{\bibfnamefont{J.~D.} \bibnamefont{Eaves}},
  \bibinfo{journal}{J. Chem. Phys.} \textbf{\bibinfo{volume}{129}},
  \bibinfo{eid}{101104} (\bibinfo{year}{2008}).

\bibitem[{\citenamefont{Jang}(2009)}]{Jang2009}
\bibinfo{author}{\bibfnamefont{S.}~\bibnamefont{Jang}}, \bibinfo{journal}{J.
  Chem. Phys.} \textbf{\bibinfo{volume}{131}}, \bibinfo{eid}{164101}
  (\bibinfo{year}{2009}).

\bibitem[{\citenamefont{Jang}(2011)}]{Jang2011}
\bibinfo{author}{\bibfnamefont{S.}~\bibnamefont{Jang}}, \bibinfo{journal}{J.
  Chem. Phys.} \textbf{\bibinfo{volume}{135}}, \bibinfo{pages}{034105}
  (\bibinfo{year}{2011}).

\bibitem[{\citenamefont{McCutcheon and Nazir}(2010)}]{McCutcheon2010}
\bibinfo{author}{\bibfnamefont{D.~P.~S.} \bibnamefont{McCutcheon}}
  \bibnamefont{and} \bibinfo{author}{\bibfnamefont{A.}~\bibnamefont{Nazir}},
  \bibinfo{journal}{New J. Phys.} \textbf{\bibinfo{volume}{12}},
  \bibinfo{pages}{113042} (\bibinfo{year}{2010}).

\bibitem[{\citenamefont{McCutcheon and
  Nazir}(2011{\natexlab{a}})}]{McCutcheon2011b}
\bibinfo{author}{\bibfnamefont{D.~P.~S.} \bibnamefont{McCutcheon}}
  \bibnamefont{and} \bibinfo{author}{\bibfnamefont{A.}~\bibnamefont{Nazir}},
  \bibinfo{journal}{Phys. Rev. B} \textbf{\bibinfo{volume}{83}},
  \bibinfo{pages}{165101} (\bibinfo{year}{2011}{\natexlab{a}}).

\bibitem[{\citenamefont{McCutcheon and
  Nazir}(2011{\natexlab{b}})}]{McCutcheon2011a}
\bibinfo{author}{\bibfnamefont{D.~P.~S.} \bibnamefont{McCutcheon}}
  \bibnamefont{and} \bibinfo{author}{\bibfnamefont{A.}~\bibnamefont{Nazir}},
  \bibinfo{journal}{J. Chem. Phys.} \textbf{\bibinfo{volume}{135}},
  \bibinfo{eid}{114501} (\bibinfo{year}{2011}{\natexlab{b}}).

\bibitem[{\citenamefont{McCutcheon et~al.}(2011)\citenamefont{McCutcheon,
  Dattani, Gauger, Lovett, and Nazir}}]{McCutcheon2011}
\bibinfo{author}{\bibfnamefont{D.~P.~S.} \bibnamefont{McCutcheon}},
  \bibinfo{author}{\bibfnamefont{N.~S.} \bibnamefont{Dattani}},
  \bibinfo{author}{\bibfnamefont{E.~M.} \bibnamefont{Gauger}},
  \bibinfo{author}{\bibfnamefont{B.~W.} \bibnamefont{Lovett}},
  \bibnamefont{and} \bibinfo{author}{\bibfnamefont{A.}~\bibnamefont{Nazir}},
  \bibinfo{journal}{Phys. Rev. B} \textbf{\bibinfo{volume}{84}},
  \bibinfo{pages}{081305} (\bibinfo{year}{2011}).

\bibitem[{\citenamefont{Silbey and Harris}(1984)}]{Silbey1984}
\bibinfo{author}{\bibfnamefont{R.}~\bibnamefont{Silbey}} \bibnamefont{and}
  \bibinfo{author}{\bibfnamefont{R.~A.} \bibnamefont{Harris}},
  \bibinfo{journal}{J. Chem. Phys.} \textbf{\bibinfo{volume}{80}},
  \bibinfo{pages}{2615} (\bibinfo{year}{1984}).

\bibitem[{\citenamefont{Harris and Silbey}(1985)}]{Harris1985}
\bibinfo{author}{\bibfnamefont{R.~A.} \bibnamefont{Harris}} \bibnamefont{and}
  \bibinfo{author}{\bibfnamefont{R.}~\bibnamefont{Silbey}},
  \bibinfo{journal}{J. Chem. Phys.} \textbf{\bibinfo{volume}{83}},
  \bibinfo{pages}{1069} (\bibinfo{year}{1985}).

\bibitem[{\citenamefont{Pach{\'o}n and Brumer}(2011)}]{pachon11}
\bibinfo{author}{\bibfnamefont{L.~A.} \bibnamefont{Pach{\'o}n}}
  \bibnamefont{and} \bibinfo{author}{\bibfnamefont{P.}~\bibnamefont{Brumer}},
  \bibinfo{journal}{J. Phys. Chem. Lett.} \textbf{\bibinfo{volume}{2}},
  \bibinfo{pages}{2728} (\bibinfo{year}{2011}).

\bibitem[{\citenamefont{Ishizaki and Fleming}(2009)}]{ishizaki09b}
\bibinfo{author}{\bibfnamefont{A.}~\bibnamefont{Ishizaki}} \bibnamefont{and}
  \bibinfo{author}{\bibfnamefont{G.~R.} \bibnamefont{Fleming}},
  \bibinfo{journal}{J. Chem. Phys.} \textbf{\bibinfo{volume}{130}},
  \bibinfo{pages}{234110} (\bibinfo{year}{2009}).

\bibitem[{\citenamefont{Carmichael}(1999)}]{carmichael1999}
\bibinfo{author}{\bibfnamefont{H.~J.} \bibnamefont{Carmichael}},
  \emph{\bibinfo{title}{Statistical Methods in Quantum Optics 1}}
  (\bibinfo{publisher}{Springer}, \bibinfo{year}{1999}).

\bibitem[{\citenamefont{Leggett et~al.}(1987)\citenamefont{Leggett,
  Chakravarty, Dorsey, Fisher, Garg, and Zwerger}}]{Leggett1987}
\bibinfo{author}{\bibfnamefont{A.~J.} \bibnamefont{Leggett}},
  \bibinfo{author}{\bibfnamefont{S.}~\bibnamefont{Chakravarty}},
  \bibinfo{author}{\bibfnamefont{A.~T.} \bibnamefont{Dorsey}},
  \bibinfo{author}{\bibfnamefont{M.~P.~A.} \bibnamefont{Fisher}},
  \bibinfo{author}{\bibfnamefont{A.}~\bibnamefont{Garg}}, \bibnamefont{and}
  \bibinfo{author}{\bibfnamefont{W.}~\bibnamefont{Zwerger}},
  \bibinfo{journal}{Rev. Mod. Phys.} \textbf{\bibinfo{volume}{59}},
  \bibinfo{pages}{1} (\bibinfo{year}{1987}).

\bibitem[{\citenamefont{Weiss}(2008)}]{weiss2008}
\bibinfo{author}{\bibfnamefont{U.}~\bibnamefont{Weiss}},
  \emph{\bibinfo{title}{Quantum Dissipative Systems}}
  (\bibinfo{publisher}{World Scientific, Singapore}, \bibinfo{year}{2008}).

\bibitem[{\citenamefont{Chin and Turlakov}(2006)}]{Chin2006}
\bibinfo{author}{\bibfnamefont{A.}~\bibnamefont{Chin}} \bibnamefont{and}
  \bibinfo{author}{\bibfnamefont{M.}~\bibnamefont{Turlakov}},
  \bibinfo{journal}{Phys. Rev. B} \textbf{\bibinfo{volume}{73}},
  \bibinfo{pages}{075311} (\bibinfo{year}{2006}).

\bibitem[{\citenamefont{Chin et~al.}(2011)\citenamefont{Chin, Prior, Huelga,
  and Plenio}}]{Chin2011}
\bibinfo{author}{\bibfnamefont{A.~W.} \bibnamefont{Chin}},
  \bibinfo{author}{\bibfnamefont{J.}~\bibnamefont{Prior}},
  \bibinfo{author}{\bibfnamefont{S.~F.} \bibnamefont{Huelga}},
  \bibnamefont{and} \bibinfo{author}{\bibfnamefont{M.~B.}
  \bibnamefont{Plenio}}, \bibinfo{journal}{Phys. Rev. Lett.}
  \textbf{\bibinfo{volume}{107}}, \bibinfo{pages}{160601}
  (\bibinfo{year}{2011}).

\bibitem[{\citenamefont{Laird et~al.}(1991)\citenamefont{Laird, Budimir, and
  Skinner}}]{Laird1991}
\bibinfo{author}{\bibfnamefont{B.~B.} \bibnamefont{Laird}},
  \bibinfo{author}{\bibfnamefont{J.}~\bibnamefont{Budimir}}, \bibnamefont{and}
  \bibinfo{author}{\bibfnamefont{J.~L.} \bibnamefont{Skinner}},
  \bibinfo{journal}{J. Chem. Phys.} \textbf{\bibinfo{volume}{94}},
  \bibinfo{pages}{4391} (\bibinfo{year}{1991}).

\bibitem[{\citenamefont{Meier and Tannor}(1999)}]{Meier1999}
\bibinfo{author}{\bibfnamefont{C.}~\bibnamefont{Meier}} \bibnamefont{and}
  \bibinfo{author}{\bibfnamefont{D.~J.} \bibnamefont{Tannor}},
  \bibinfo{journal}{J. Chem. Phys.} \textbf{\bibinfo{volume}{111}},
  \bibinfo{pages}{3365} (\bibinfo{year}{1999}).

\bibitem[{\citenamefont{Geva et~al.}(2000)\citenamefont{Geva, Rosenman, and
  Tannor}}]{Geva2000}
\bibinfo{author}{\bibfnamefont{E.}~\bibnamefont{Geva}},
  \bibinfo{author}{\bibfnamefont{E.}~\bibnamefont{Rosenman}}, \bibnamefont{and}
  \bibinfo{author}{\bibfnamefont{D.}~\bibnamefont{Tannor}},
  \bibinfo{journal}{J. Chem. Phys.} \textbf{\bibinfo{volume}{113}},
  \bibinfo{pages}{1380} (\bibinfo{year}{2000}).

\bibitem[{\citenamefont{Chandler}(1991)}]{chandler1991}
\bibinfo{author}{\bibfnamefont{D.}~\bibnamefont{Chandler}}, in
  \emph{\bibinfo{booktitle}{Liquids, Freezing and the Glass Transition, Les
  Houches}}, edited by
  \bibinfo{editor}{\bibfnamefont{D.}~\bibnamefont{Levesque}},
  \bibinfo{editor}{\bibfnamefont{J.~P.} \bibnamefont{Hansen}},
  \bibnamefont{and}
  \bibinfo{editor}{\bibfnamefont{J.}~\bibnamefont{Zinn-Justin}}
  (\bibinfo{publisher}{Elsevier}, \bibinfo{address}{New York},
  \bibinfo{year}{1991}), p. \bibinfo{pages}{193}.

\bibitem[{\citenamefont{Feynman and Vernon}(1963)}]{Feynman63}
\bibinfo{author}{\bibfnamefont{R.~P.} \bibnamefont{Feynman}} \bibnamefont{and}
  \bibinfo{author}{\bibfnamefont{F.~L.} \bibnamefont{Vernon}},
  \bibinfo{journal}{Ann. Phys. (NY)} \textbf{\bibinfo{volume}{24}},
  \bibinfo{pages}{118} (\bibinfo{year}{1963}).

\bibitem[{\citenamefont{Schulman}(1986)}]{schul86}
\bibinfo{author}{\bibfnamefont{L.~S.} \bibnamefont{Schulman}},
  \emph{\bibinfo{title}{Techniques and Applications of Path Integration}}
  (\bibinfo{publisher}{Wiley}, \bibinfo{address}{New York},
  \bibinfo{year}{1986}).

\bibitem[{\citenamefont{Grabert et~al.}(1988)\citenamefont{Grabert, Schramm,
  and Ingold}}]{Ingold88}
\bibinfo{author}{\bibfnamefont{H.}~\bibnamefont{Grabert}},
  \bibinfo{author}{\bibfnamefont{P.}~\bibnamefont{Schramm}}, \bibnamefont{and}
  \bibinfo{author}{\bibfnamefont{G.-L.} \bibnamefont{Ingold}},
  \bibinfo{journal}{Phys. Rep.} \textbf{\bibinfo{volume}{168}},
  \bibinfo{pages}{115} (\bibinfo{year}{1988}).

\bibitem[{\citenamefont{Kleinert}(2004)}]{kleinert04}
\bibinfo{author}{\bibfnamefont{H.}~\bibnamefont{Kleinert}},
  \emph{\bibinfo{title}{Path Integrals in Quantum Mechanics, Statistics, and
  Polymer Physics, and Financial Markets}} (\bibinfo{publisher}{World
  Scientific}, \bibinfo{address}{Singapore}, \bibinfo{year}{2004}),
  \bibinfo{edition}{3rd} ed.

\bibitem[{\citenamefont{Moix et~al.}(2012)\citenamefont{Moix, Zhao, and
  Cao}}]{jeremy2012}
\bibinfo{author}{\bibfnamefont{J.}~\bibnamefont{Moix}},
  \bibinfo{author}{\bibfnamefont{Y.}~\bibnamefont{Zhao}}, \bibnamefont{and}
  \bibinfo{author}{\bibfnamefont{J.}~\bibnamefont{Cao}},
  \bibinfo{journal}{submitted}  (\bibinfo{year}{2012}).

\end{thebibliography}
\end{document}